\documentclass[conference]{IEEEtran}
\pdfoutput=1

\usepackage[utf8]{inputenc}
\usepackage[T1]{fontenc}
\usepackage{microtype}

\usepackage{amsmath,amssymb,amsfonts}
\usepackage{siunitx}

\usepackage{array}
\usepackage{booktabs}
\usepackage[final]{graphicx}
\usepackage{flafter}            
\usepackage[section]{placeins}  
\usepackage{float}              

\usepackage{capt-of}
\usepackage{xcolor}
\usepackage{csquotes}
\usepackage{tikz}

\usepackage{algorithm}
\usepackage[noend]{algpseudocode}

\usepackage[hidelinks]{hyperref}
\usepackage[capitalize,noabbrev]{cleveref}
\crefname{figure}{Fig.}{Figs.}
\Crefname{figure}{Fig.}{Figs.}
\crefname{table}{Table}{Tables}
\Crefname{table}{Table}{Tables}

\usepackage{balance}

\DeclareGraphicsExtensions{.pdf,.png,.jpg}
\graphicspath{{figs/}}

\newcommand{\phide}{\phi_{\mathrm{DE}}}
\newcommand{\phihz}{\phi_{\mathrm{HZ}}}
\newcommand{\kap}{\kappa}
\newcommand{\U}{U}
\DeclareMathOperator{\E}{\mathbb{E}}

\newcommand{\figw}{0.95\linewidth}

\newif\ifanonymize
\anonymizefalse

\ifanonymize
  \title{X\textendash SYCON: Xylem\textendash Inspired Passive Gradient Control for Communication\textendash Free Swarm Response in Dynamic Disaster Environments}
  \author{\IEEEauthorblockN{Anonymous Authors}\IEEEauthorblockA{}}
\else
  \title{X\textendash SYCON: Xylem\textendash Inspired Passive Gradient Control for Communication\textendash Free Swarm Response in Dynamic Disaster Environments}
  \author{
    \IEEEauthorblockN{Arthur Ji Sung Baek}
    \IEEEauthorblockA{Independent Researcher\\São Paulo, SP, Brazil\\
    \href{mailto:arthurbaek08@gmail.com}{arthurbaek08@gmail.com}}
    \and
    \IEEEauthorblockN{Geoffrey Martin}
    \IEEEauthorblockA{Cornell University\\Ithaca, NY, USA\\
    \href{mailto:ghm58@cornell.edu}{ghm58@cornell.edu}}
  }
\fi

\begin{document}
\maketitle
\thispagestyle{plain}
\pagestyle{plain}

\begin{abstract}
We present X--SYCON, a xylem-inspired multi-agent architecture in which coordination emerges from passive field dynamics rather than explicit planning or communication. Incidents (demands) and obstructions (hazards) continually write diffusing and decaying scalar fields, and agents greedily ascend a local utility \( U = \phi_{\mathrm{DE}} - \kappa \,\phi_{\mathrm{HZ}} \) with light anti-congestion and separation. A beaconing rule triggered on first contact temporarily deepens the local demand sink, accelerating completion without reducing time-to-first-response. Across dynamic, partially blocked simulated environments, we observe low miss rates and stable throughput with interpretable, tunable trade-offs over carrier count, arrival rate, hazard density, and hazard sensitivity \(\kappa\). We derive that a characteristic hydraulic length scale predicts recruitment range in a continuum approximation, and we provide a work-conservation (Ohm-law) bound consistent with sublinear capacity scaling with team size. Empirically: (i) soft hazard penalties yield fewer misses when obstacles already block motion; (ii) throughput saturates sublinearly with carriers while reliability improves sharply; (iii) stronger arrivals can reduce misses by sustaining sinks that recruit help; and (iv) phase-stability regions shrink with hazard density but are recovered by more carriers or higher arrivals. We refer to X--SYCON as an instance of Distributed Passive Computation and Control, and we evaluate it in simulations modeling communication-denied disaster response and other constrained sensing--action regimes.
\end{abstract}

\section{Introduction}
Major earthquakes, structural collapses, and industrial accidents routinely produce environments that defeat the assumptions of modern multi–robot systems: radio links are intermittent or denied, line–of–sight sensing is blocked by debris and dust, maps become stale minutes after they are made, and the cost of waiting for a global plan can be measured in lives \cite{Murphy2014}. In such settings, classical multi–agent machinery—shared situational awareness, explicit task assignment, and multi–hop coordination—either fails outright or introduces fragile single points of failure. This work asks: \emph{To what extent can multi-robot coordination be achieved without inter-robot communication or centralized tasking?}

Plant xylem provides an analog for passive, distributed transport, which motivates our approach \cite{TyreeZimmermann2002,Sperry2000}. Xylem transports water from roots to leaves via passive physics: a potential difference drives flow through a resistance network; no cell “plans,” yet the whole organism self–organizes a stable, robust transport process \cite{TyreeZimmermann2002,Sperry2000}. We translated this principle into swarm robotics. In our architecture, \textbf{X--SYCON}, incidents (demands) and obstructions (hazards) continually write scalar fields that \emph{diffuse and decay} across the environment. Each robot (“carrier”) moves by greedily ascending the \emph{local} utility
\[
\U \;=\; \phide \;-\; \kap\,\phihz,
\]
with light anti–congestion and separation. There is no explicit planning or messaging; coordination is achieved via diffusion, decay, and local reseeding, implementing a physical prioritization and routing substrate. A simple \emph{beaconing} rule—activated on first contact with a demand—transiently deepens the local sink, recruiting nearby carriers to finish faster without reducing time–to–first–response.

This perspective differs in important ways from well–known bio–inspired approaches. Classical potential–field robotics \cite{Khatib1985} typically assumes known targets and static, designer–chosen potentials; pheromone and stigmergic systems \cite{Payton2005} rely on agents to \emph{deposit} messages in the world and tune interaction protocols; market–based methods exchange bids or roles over a communication fabric; frontier exploration and SLAM assume that maps can be maintained. X--SYCON, in contrast, treats the environment itself as a \emph{passive analog computer}: demands and hazards reseed fields that obey simple transport laws; agents only sense and climb local gradients. Information is implicitly conveyed through diffusive field dynamics. Consequently, the architecture is (i) communication–free by design, (ii) inherently distributed and local, and (iii) interpretable in the language of transport theory.

Two theoretical lenses structure the design. First, the recruitment range is controlled by a \emph{hydraulic length scale} \(\ell \approx \sqrt{D/\lambda}\) that arises from the continuum approximation \(\partial_t \phi \approx D\nabla^2\phi - \lambda\phi + S\). Tuning \(\ell\) balances exploration (longer reach) against clumping (shorter reach), providing a physical knob for coverage versus responsiveness. Second, an \emph{Ohm–law service bound} \(\mu_{\max} \lesssim C\cdot\text{service-rate}/\mathbb{E}[B]\) links throughput to carrier count \(C\) and per–task service budget \(B\). This bound is consistent with sublinear capacity scaling with team size, even when reliability improves sharply—an effect we verify empirically. Together, these notions allow us to talk about coordination performance without invoking global plans: diffusion sets who can “hear” a task; resistance (hazard penalty \(\kap\)) sets which corridors are preferred; and the network’s effective conductance rises transiently when multiple carriers recruit to a beaconed sink.

We evaluated in settings motivated by disaster response. Hazards (blocked or unsafe cells) dynamically appear and disappear; paths open and close; demand spawns stochastically; energy is precious; and reliability matters at least as much as the raw throughput. Under these conditions, two design choices are crucial. First, hazards already impose \emph{hard} constraints via blocked cells; therefore, over–penalizing their diffusive halos can reduce recruitment across safe corridors, and a large \( \kappa \) suppresses recruitment so strongly that sinks fail to pull help across safe corridors. In our simulations, smaller non-zero \( \kappa \) values are associated with lower miss rates; \( \kappa \) acts as a tunable trade-off parameter. Second, the beaconing rule is configured to collapse the tail of the \emph{completion latency (post–contact)} while leaving the time-to-first-response unchanged, by temporarily increasing the local demand source and recruiting nearby carriers to finish faster. Empirically, this manifests as a reduction in \emph{Time in System (TiS)} at comparable arrivals, whereas TTFR remains unchanged.

From an engineering standpoint, X--SYCON aims to be deployable on modest platforms. Per–agent computation is constant–time local sensing and an argmax over neighbors; the memory per agent is \(O(1)\); there is no reliance on synchronized clocks, shared coordinate frames, or reliable radio. The field machinery can be realized in simulation (our 33\(\times\)33 proxy world) or approximated in hardware via on–board filtering of broadcast–free measurements (e.g., gas concentration, acoustic intensity, and light) that naturally diffuse and decay. This approach may scale to larger collectives because it relies on local sensing and passive physical fields.

Empirically, we evaluated the architecture with \(\,2{,}560\) BehaviorSpace runs spanning carrier count, arrival rate, hazard density, and hazard sensitivity. The data reveal three recurring patterns, which suggest potential design considerations: (i) penalize hazards \emph{lightly} when obstacles are hard; (ii) throughput saturates sublinearly with team size while reliability improves dramatically; and (iii) stronger arrivals can \emph{reduce} misses because sustained sinks recruit help more often, especially with beaconing. We also chart phase–stability regions (fractions of runs below a miss–rate threshold) that shrink as hazards intensify but are recoverable by increasing carriers or arrivals. These patterns are consistent with a transport interpretation in which diffusion–decay influences recruitment reach, hazard penalty affects path selection, and beaconing increases the local effective attraction.

\subsection*{Contributions}
We make three contributions:
\begin{enumerate}
  \item \textbf{Passive–gradient coordination without communication.} We introduce X--SYCON, a xylem–inspired architecture in which demands and hazards write diffusing, decaying fields and agents climb a local utility \( U = \phi_{\mathrm{DE}} - \kappa\,\phi_{\mathrm{HZ}} \). In a dynamic, partially blocked world, this yields low miss rates and stable throughput without messaging or task assignment; beaconing reduces the Time in System (creation$\to$completion/miss) while preserving the time–to–first–response.
  \item \textbf{Distributed Passive Computation: a unifying lens.} We demonstrate that diffusion, decay, urgency, and beaconing implement priority, load–sharing, and routing in physics. A hydraulic length scale \(\ell \approx \sqrt{D/\lambda}\) relates to the recruitment range and exploration–clumping behavior, and an Ohm–law bound is consistent with the sublinear capacity scaling with carriers.
  \item \textbf{Policy–level trade–offs that are quantified and tunable.} We mapped the energy–reliability frontier and phase–stability regions across hazards, carrier counts, and arrivals. Designers can tune \(\kappa\), \(C\), and \(p_{\mathrm{new}}\) to select operating points that meet power budgets and reliability targets, with behavior that generalizes across environments.
\end{enumerate}

Evaluated under communication-denied, these results characterize a class we refer to as \emph{Distributed Passive Computation and Control}: swarms coordinated by physical processes that neither centrally compute nor explicitly communicate, suited to the messy realities of communication–denied response and other constrained sensing–action regimes.

\subsection*{Problem Statement and Assumptions}
\textbf{Problem.} We study the decentralized service of stochastically arriving demands in a dynamic, partially blocked environment, \emph{without} communication or centralized tasking. Each carrier must decide motion and service using only local samples of the demand and hazard fields, seeking to minimize missed tasks and \emph{Time in System (TiS)} under energy constraints while respecting hazard avoidance.
    
\textbf{Assumptions.}
\begin{itemize}
    \item[(i)] \textbf{Local sensing:} each carrier accurately samples $(\phide,\phihz)$ and neighbor occupancy within its von~Neumann neighborhood of each tick.
    \item[(ii)] \textbf{Synchronized updates:} diffusion/decay, motion, and service occur on a common tick timescale; carriers do not share clocks cycles beyond this discrete step.
    \item[(iii)] \textbf{Hazards:} blocked cells are hard constraints (no traversal); the diffusive hazard halo $\phihz$ encodes \emph{soft} avoidance via utility penalty $\kap$.
    \item[(iv)] \textbf{No messaging map:} agents do not broadcast messages, maintain global maps, or exchange roles; policies are identical and purely local.
    \item[(v)] \textbf{Lightweight compute:} per-agent computation and memory are $O(1)$; action selection is an argmax over four neighbors using $\,\U=\phide-\kap\,\phihz\,$ with light anti-congestion and tie-breaking noise.
\end{itemize}

These assumptions bound the scope of target—communication-denied, rapidly changing environments—where coordination must emerge from passive field dynamics (diffusion, decay, and reseeding) rather than explicit signaling.

\section{Related Work}

\subsection{From artificial potentials to field-mediated swarms}
Artificial potential fields have been a workhorse for local navigation since Khatib’s seminal work \cite{Khatib1985}, with well-known drawbacks such as local minima and oscillations in clutter \cite{KorenBorenstein1991}. Coverage and formation control recast these ideas in distributed optimization: gradient-based deployment and centroidal Voronoi control for sensor networks \cite{Howard2002,Cortes2004}, virtual leaders and flocking for coordination \cite{LeonardFiorelli2001,OlfatiSaber2006}, and motion-planning perspectives that highlight explicit goal specification and global search \cite{LaValle2006,Latombe1991}. 
Compared with artificial potential fields, X--SYCON uses fields written by incidents and obstacles and relies solely on local sensing. Agents do not communicate goals or plans; instead, the environment \emph{computes} soft priorities by diffusing and decaying fields. Our utility $U=\phide-\kap\phihz$ looks superficially like an artificial potential, but its sources are generated by incidents and obstacles rather than by a central planner. The empirical result that a \emph{small but non-zero} hazard penalty improves reliability (\cref{fig:res-kappa}) addresses the classic failure mode of potential fields in clutter: over-penalizing hazard halos suppresses recruitment and amplifies dead zones.

\begin{figure}[t]
  \centering
  \includegraphics[width=88mm]{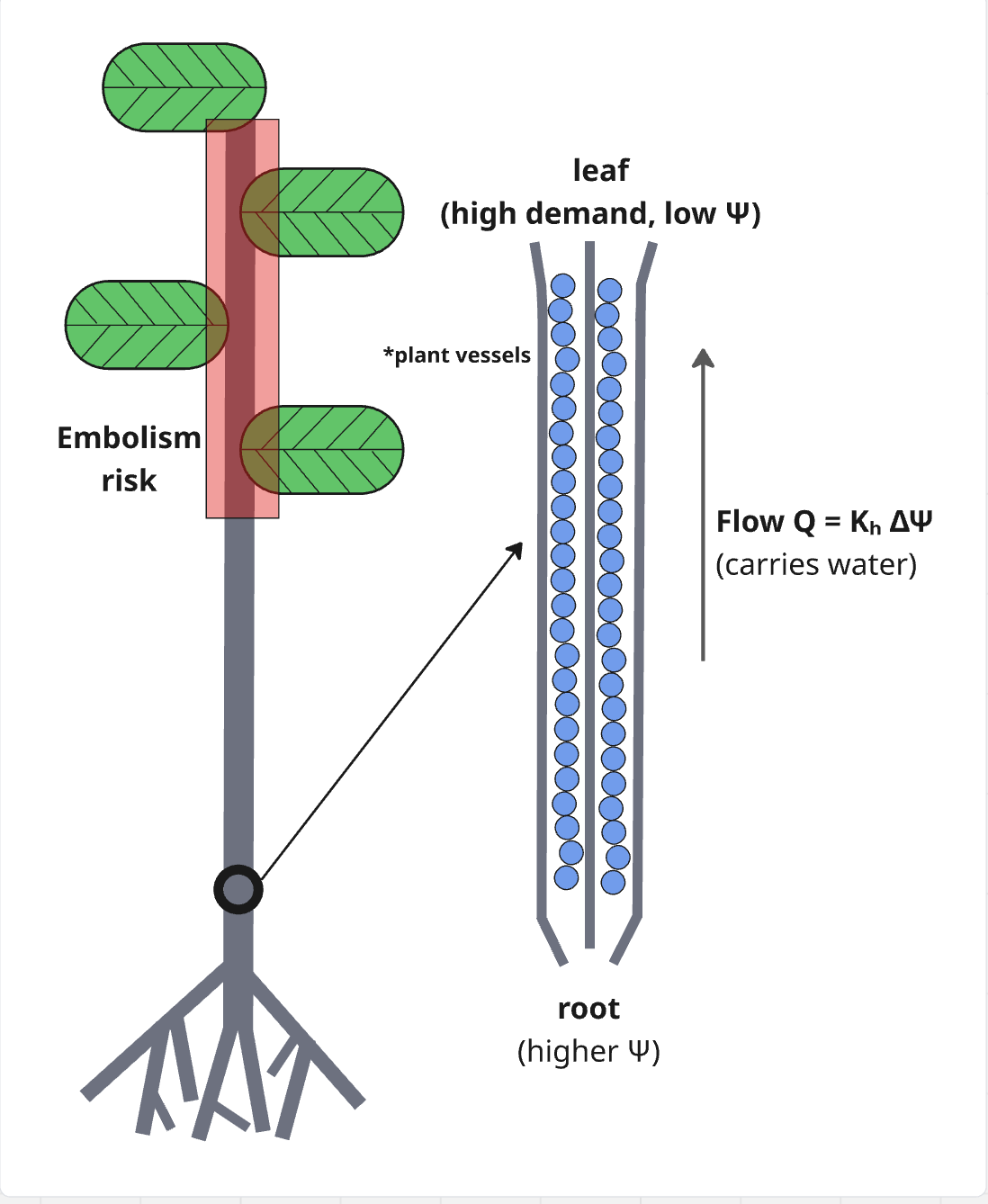}
  \caption{%
  \textbf{Botany (xylem).} Water flows upward from roots (higher $\Psi$) to leaves
  (low $\Psi$) driven by a potential drop $\Delta\Psi$, with Ohm-like relation
  $Q=K_h\,\Delta\Psi$. The vessel bundle (right) acts as a parallel conduit;
  conductance $K_h$ reflects lumen geometry and connectivity. Near distal
  branchings the water column is under high tension, increasing \emph{embolism
  risk} (red zone).}
  \label{fig:botany-xylem}
\end{figure}

\begin{figure}[t]
  \centering
  \includegraphics[width=\columnwidth]{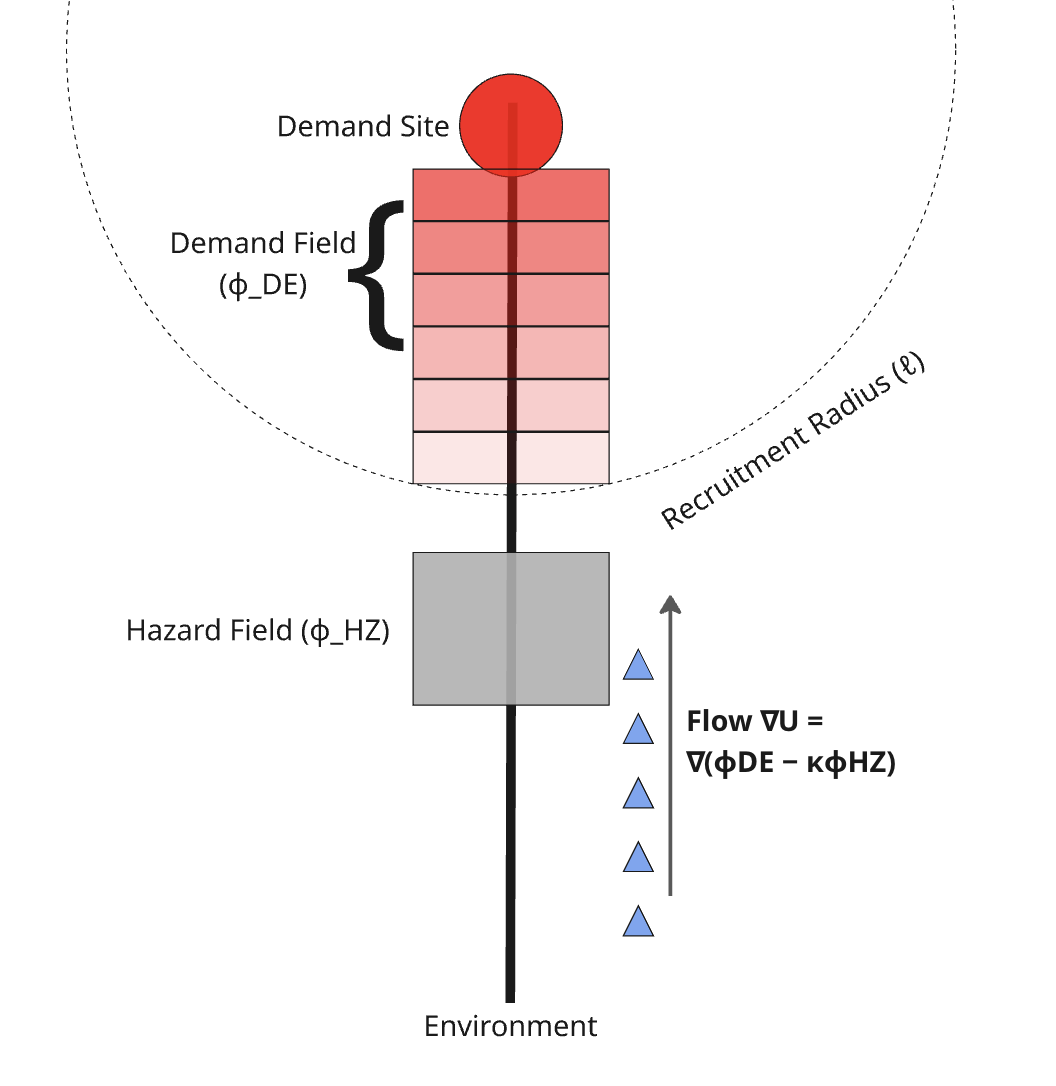}
  \caption{\textbf{X--SYCON (passive fields) schematic.}
  The red circle marks the demand site; the stacked red bars depict the demand field $\phi_{\mathrm{DE}}$ that weakens with distance (a sink). The dashed arc indicates the recruitment radius $\ell$, within which the sink effectively attracts carriers. A gray block depicts a hazard region (hazard field $\phi_{\mathrm{HZ}}$) that imposes a soft resistance. Blue triangles illustrate carriers moving up the net utility gradient, following the motion rule $\nabla U = \nabla(\phi_{\mathrm{DE}} - \kappa \phi_{\mathrm{HZ}})$ toward the demand while bending around hazards. This panel is a schematic—not a map—intended to show how demand attraction, hazard penalty, and recruitment radius jointly shape carrier flow.}
  \label{fig:xsycon-schematic}
\end{figure}

\subsection{Swarm robotics, stigmergy, and “digital pheromones”}
\begin{figure}[t]
  \centering
  \includegraphics[width=88mm]{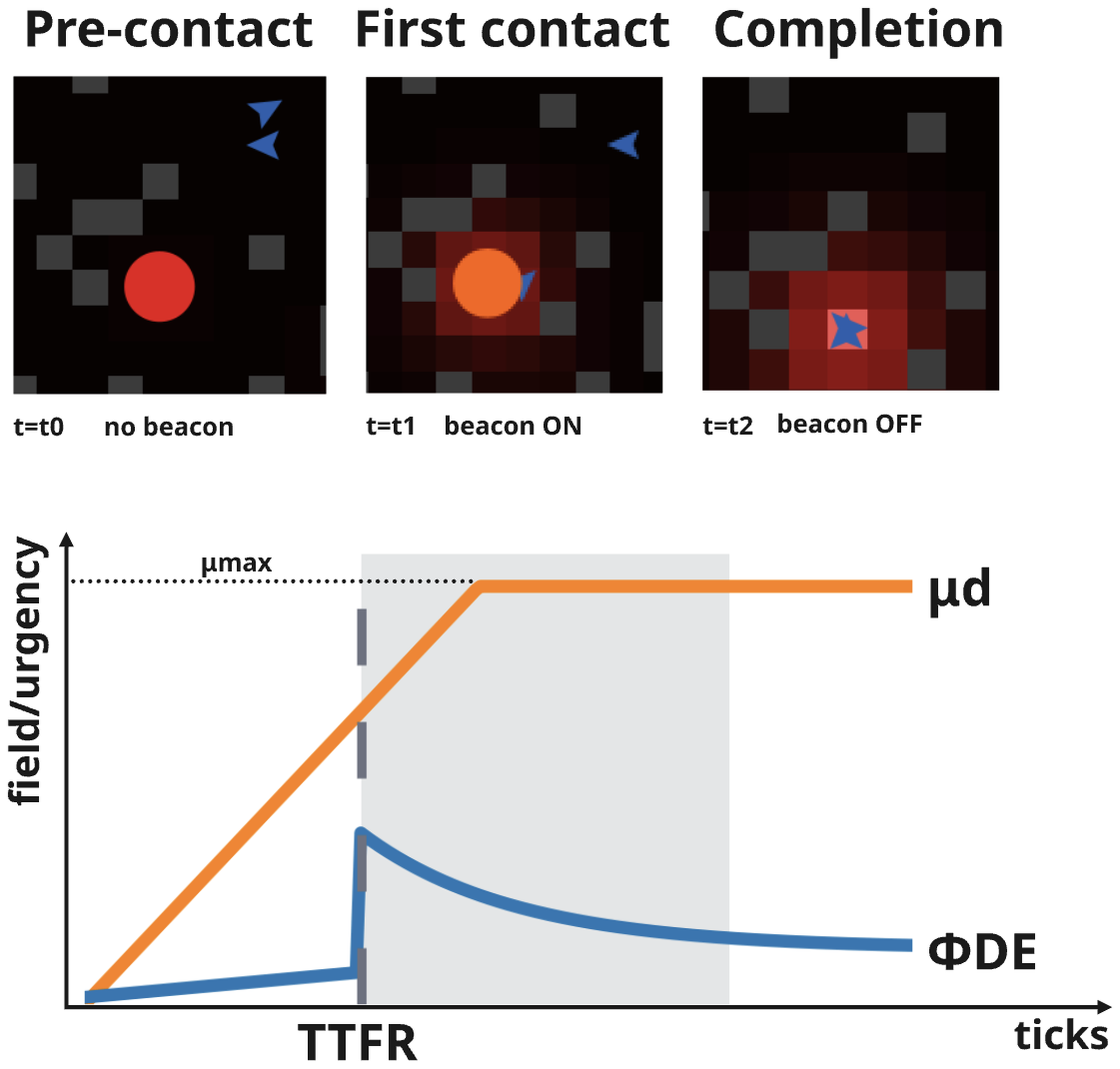}
  \caption{\textbf{Beaconing timeline and effect on completion latency.}
  Top row: \emph{Pre-contact} ($t=t_0$) shows carriers (blue) ascending the demand field $\phi_{\mathrm{DE}}$ around obstacles (dark gray); the demand site is at the center. \emph{First contact} ($t=t_1$) switches a local beacon (\emph{beacon-bonus} in \eqref{eq:sde}), deepening $\phi_{\mathrm{DE}}$ and bending nearby flows. \emph{Completion} ($t=t_2$) turns the beacon off and the field relaxes. Bottom: the orange curve is task urgency $u_d$ rising linearly and clamping at $u_{\max}$; the blue curve is the local $\phi_{\mathrm{DE}}$ at the demand site, which jumps at $t_1$ due to beaconing and then decays as service proceeds. ... The dashed marker at $t_1$ denotes time-to-first-response (TTFR), which is unchanged by construction. In our implementation we report \emph{time in system} (creation$\to$completion/miss); completion latency ($t_2{-}t_1$) is qualitatively reduced by beaconing when first contact occurs.}

  \label{fig:beaconing}
\end{figure}

Swarm robotics has long leveraged local rules and indirect coordination to achieve scalable behaviors \cite{Brambilla2013,Sahin2005}. Stigmergy—in which agents modify and sense a shared medium—provides a general mechanism for coordination without messaging \cite{MataricWerger1997,BeckersHollandDeneubourg2000}. Robotic “pheromone” systems demonstrate how deposited fields guide dispersion, foraging, and task allocation \cite{Payton2005,SvennebringKoenig2004,Campo2007}. 
X--SYCON is similar to stigmergic approaches in using a shared medium sensed locally, but differs in two key respects. First, the \emph{two-field} structure (demand and hazard) implements a vector of priorities rather than a single scalar trace, thereby providing a tunable trade-off between recruitment and risk. Second, our \emph{beaconing} rule (first-contact local reseed) shortens the post-contact completion interval without changing the time-to-first-response (\cref{fig:beaconing}), which is distinct from pheromone amplification, which typically accelerates both approach and completion. The measured coverage entropy (\cref{fig:res-entropy}) also situates X--SYCON between uniform exploration and highly clustered behavior, complementing classic dispersion/foraging metrics.

\subsection{Reaction--diffusion, chemotaxis, and computation with physics}
The mathematical basis for field-mediated behaviors dates to reaction--diffusion and chemotaxis models \cite{Turing1952,KellerSegel1971,Murray2002}. These PDEs encode the local production, decay, and diffusion shape macroscopic patterns, an idea exploited in unconventional computing \cite{Adamatzky2009}, and the control of taxis-like responses \cite{HillenPainter2009}. Our derivation of a hydraulic length scale $\ell\approx\sqrt{D/\lambda}$ (Methods) is a standard Green’s-function result in such media, but here it provides design guidance: $\ell$ sets a recruitment radius and, hence, the exploration--clumping balance. In contrast to controllers that \emph{sample} fields produced by onboard computation, X--SYCON allows the environment’s physics to perform the aggregation and attenuation, which encode task urgency and obstacle risk. We use the term \emph{Distributed Passive Computation and Control} to describe this reliance on the physical field dynamics.

\subsection{Plant hydraulics as design inspiration}
\label{subsec:xylem}
Xylem transports water down a potential gradient $\Delta\Psi$ generated by transpiration, with the flow obeying an Ohm--law analogue $Q = K_h \Delta\Psi$ (\cref{fig:botany-xylem}) where conductance $K_h$ reflects vessel geometry and embolism risk \cite{TyreeZimmermann2002,Nobel2009,TaizZeiger2014}. The system is passive, distributed, and local: there are no cell plans, yet the plant flow self--organizes \cite{Sperry2000}. We map this directly in X--SYCON: demands act as tunable \emph{sinks} reseeding $\phide$; obstacles reseed hazards $\phihz$ (\cref{fig:xsycon-schematic}); and in settings with uncrossable obstacles, smaller nonzero \(\kappa\) values were associated with lower miss rates (\cref{fig:res-kappa}). The contraction of stable phases as hazard density increases and their recovery by stronger sinks or more carriers (\cref{fig:res-phase}) echoes conductance--demand tradeoffs in hydraulic networks.

\subsection{Disaster robotics and communication-denied autonomy}
Urban search-and-rescue (USAR) imposes tight constraints on communication, energy, and time \cite{Murphy2014}. Recent systems advances (e.g., the DARPA Subterranean Challenge) emphasize robustness in GPS/Comms-denied settings through autonomy and environment-aware sensing \cite{Orekhov2023,Tranzatto2022,Ebadi2024}. Typical strategies rely on SLAM, frontier exploration, and team-level task allocation, all of which presuppose nontrivial communication or shared maps. X--SYCON targets a distinct regime: when communication is intermittent or absent and task discovery is highly local, we can still obtain low miss rates and stable throughput by exploiting passive field dynamics. The sublinear saturation of throughput with carrier count (\cref{fig:res-throughput}) and phase-stability maps across hazard densities (\cref{fig:res-phase}) provide quantitative design charts for sizing teams and picking arrival rates without global coordination.

\subsection{Hardware exemplars and scale considerations}
Large collectives such as Kilobots show that minimal agents can achieve sophisticated group behaviors under severe resource constraints \cite{Rubenstein2014}, whereas termite-inspired builders demonstrate construction with purely local rules \cite{Werfel2014Science}. Animal groups further illustrate how simple interaction laws yield robust swarming and lane formation \cite{Reynolds1987,Couzin2005}. X--SYCON’s reliance on the local sampling of broadcast-like fields (e.g., environmental markers, radio beacons, thermal/chemical cues, or pre-computed maps turned into virtual fields) may be implementable with similar hardware capabilities. Our energy--reliability frontier (\cref{fig:res-energy}) allows explicit comparison under power constraints: designers can dial reliability against motion/hover energy without a communication channel. For reproducibility and parameter sweeps we used NetLogo/BehaviorSpace \cite{WilenskyRand2015,RailsbackGrimm2019}; the same logging pipeline (service throughput, miss rate, TiS, and coverage entropy \cite{Shannon1948}) can be mirrored on embedded platforms.

\subsection{Positioning}
In summary, X--SYCON differs from classical potential-field navigation by allowing \emph{incidents and obstacles} to write the potentials locally; from swarm task-allocation work by avoiding messaging and role negotiation \cite{GerkeyMataric2004}; and from pheromone robotics by using a two-field passive computation that separates demand attraction from hazard repulsion with a tunable coupling $\kap$. In our simulations, we observed: (i) low misses at modest $\kap$, (ii) sublinear but stable throughput scaling, and (iii) a tunable energy--reliability frontier that remains consistent across hazard regimes (\cref{fig:res-kappa,fig:res-throughput,fig:res-energy,fig:res-phase}). 

\section{Methods}
\label{sec:methods}

\subsection{World, Fields, and Events}
We simulate a $33{\times}33$ lattice with von Neumann neighborhoods (no diagonals). Each patch stores demand and hazard fields, $\phide,\phihz\in\mathbb{R}_{\ge 0}$, and a Boolean obstruction. Time is discrete (ticks). At each tick,

\begin{align}
\phide^{t+1}(x)
  &= (1-\delta_{\mathrm{DE}})\,\phide^t(x) - \beta_{\mathrm{DE}} + S_{\mathrm{DE}}(x,t) \notag\\
  &\quad + \mathrm{Diffuse}\!\big(\phide^t; D_{\mathrm{DE}}\big) \label{eq:de}\\[2pt]
\phihz^{t+1}(x)
  &= (1-\delta_{\mathrm{HZ}})\,\phihz^t(x) + S_{\mathrm{HZ}}(x,t) \notag\\
  &\quad + \mathrm{Diffuse}\!\big(\phihz^t; D_{\mathrm{HZ}}\big). \label{eq:hz}
\end{align}

\paragraph*{Update order (operator split)}
We use an operator-split step for each tick: (i) apply proportional decay and the absolute leak on $\phide$, (ii) apply sources $S_{\mathrm{HZ}}$ (hazard reseed) and $S_{\mathrm{DE}}$ (demand reseed with capping), and then (iii) apply NetLogo's conservative von~Neumann diffusion to both fields. This matches the order of implementation.

Here $\delta_{\mathrm{DE}},\delta_{\mathrm{HZ}}\in[0,1)$ are proportional decays, $\beta_{\mathrm{DE}}\ge 0$ is an absolute leak on $\phide$, and $\mathrm{Diffuse}(\cdot;D)$ is NetLogo’s conservative von~Neumann diffusion with coefficient $D$. Hazards reseed on obstructed patches as $S_{\mathrm{HZ}}(x,t)=12\,\mathbb{I}\{\text{blocked}(x)\}$. Demands contribute to $\phide$ via $S_{\mathrm{DE}}$ at their locations.

\begin{figure}[!t]
  \centering
  \includegraphics[width=\figw]{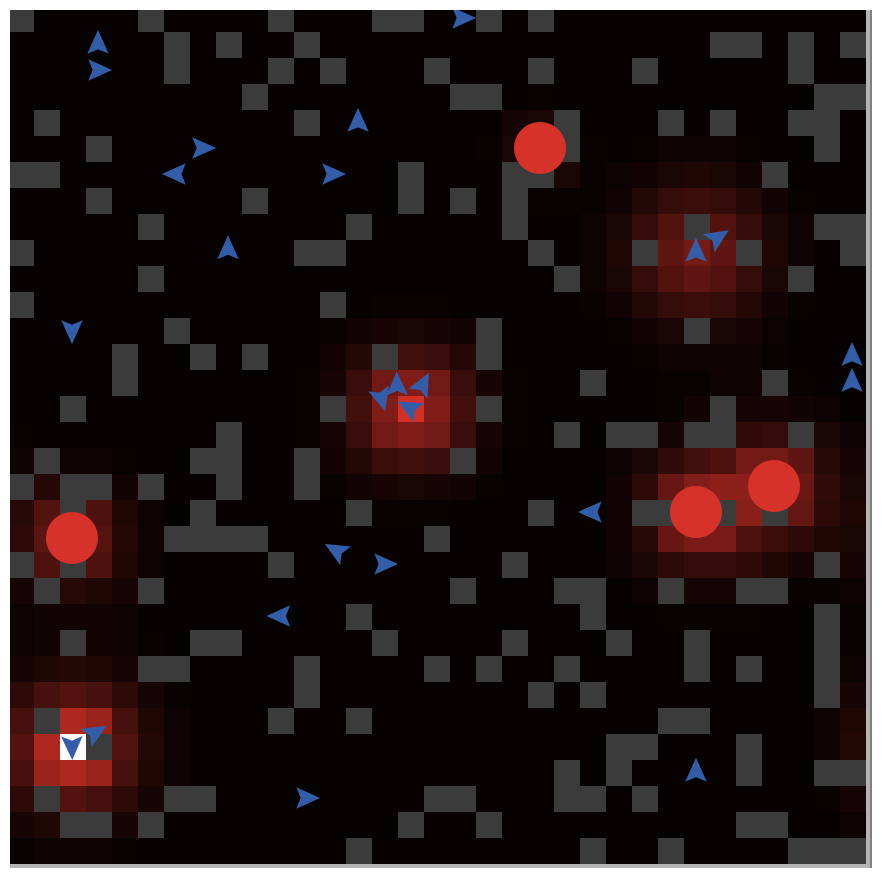}
  \caption{Qualitative snapshot of the simulated world.
  Blue triangles: carriers; red circles: demands; dark gray squares: blocked
  cells (hazard sources). Red glow indicates the demand field $\phi_{\mathrm{DE}}$.}
  \label{fig:qual-snapshot}
\end{figure}

\paragraph*{Dynamic hazards}
The obstruction fraction mean-reverts toward a target $f^\star\in[0,1]$. With probability $p_{\mathrm{hazard\text{-}change}}$ per tick, we flip $\approx 5\%$ of the gap toward $f^\star$.

\begin{figure}[!t]
  \centering
  \includegraphics[width=\columnwidth]{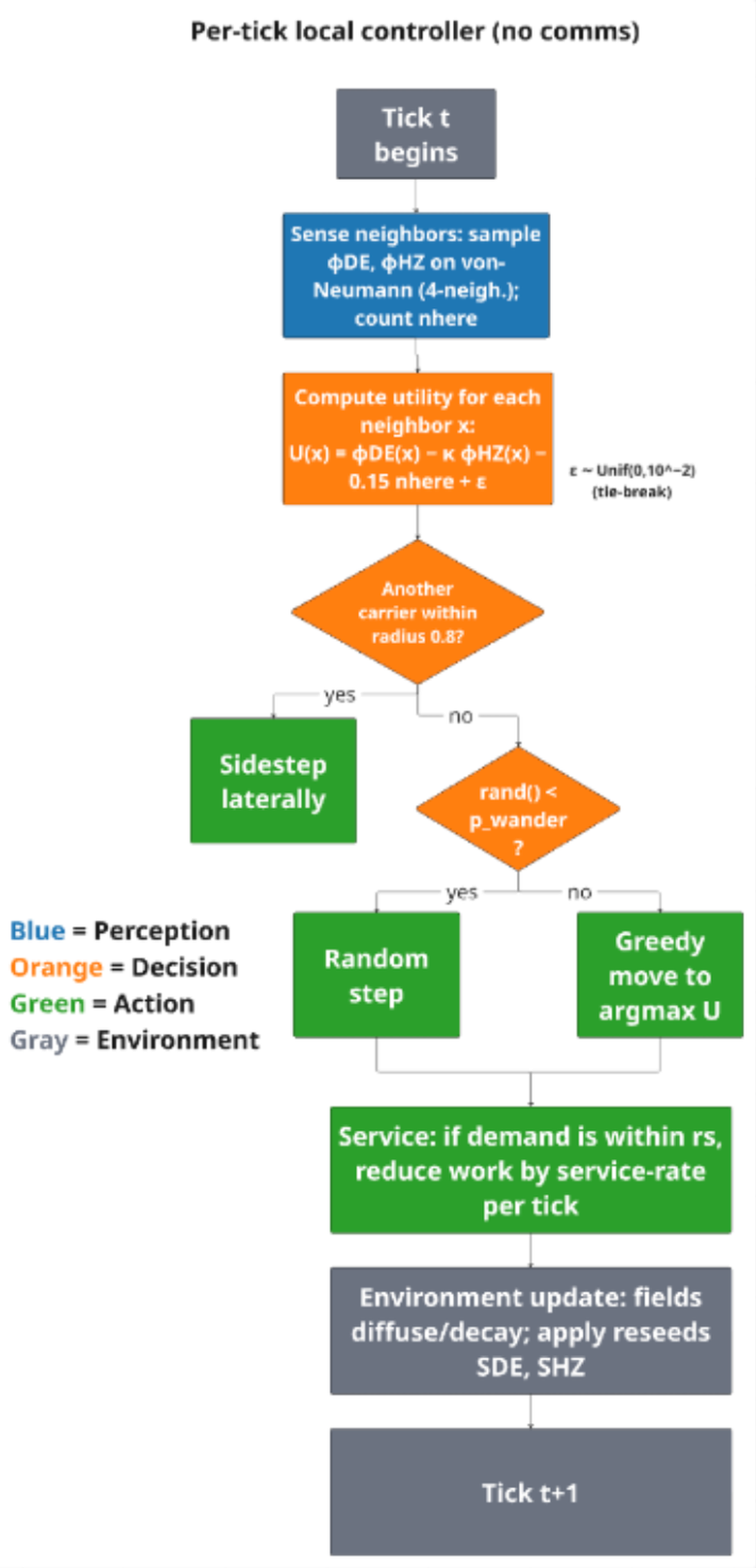}
  \vspace{2pt}
  \caption{Per-tick local controller: sense $\phi_{DE}, \phi_{HZ}$; compute $U$; separation/wander checks; random/greedy step; service; environment diffuses/decays. No communication.}
  \label{fig:controller}
\end{figure}

\paragraph*{Units and normalization}
All quantities are expressed in normalized simulation units. One tick advances both field transport and carrier motion; TTFR and other times have been reported in ticks. The lattice spacing is one spatial unit (one carrier step). The field values $(\phide,\phihz)$ and utility $\U$ are dimensionless. In our runs $\phide$ is capped by \texttt{de-cap} (=25) and $\phihz$ reseeds to 12 on blocked patches; the controller uses local differences, making the absolute scale less critical for control than \emph{relative} gradients (\cref{fig:controller}) (If desired for visualization, a normalized field $\tilde\phi=\phi/\texttt{de-cap}$ can be reported).

\begin{table}[!t]
  \centering\footnotesize
  \setlength{\tabcolsep}{6pt}\renewcommand{\arraystretch}{1.05}
  \caption{Legend}
  \label{tab:notation}
  \begin{tabular}{@{}ll@{}}
    \toprule
    Symbol & Meaning \\
    \midrule
    $\phide,\,\phihz$ & Demand / hazard fields \\
    $D_{\mathrm{DE}},\,D_{\mathrm{HZ}}$ & Diffusion coefficients \\
    $\delta_{\mathrm{DE}},\,\delta_{\mathrm{HZ}}$ & Proportional decays \\
    $\beta_{\mathrm{DE}}$ & Absolute leak on $\phide$ (per tick) \\
    $\ell$ & Hydraulic length scale $\approx\sqrt{D/\lambda}$ \\
    $\kap$ & Hazard sensitivity penalty in $\U$ \\
    $r_s$ & Service radius \\
    $C$ & Number of carriers \\
    $p_{\mathrm{new}}$ & Demand arrival probability \\
    $\mu_{\max}$ & Service bound (Eq.~\eqref{eq:bound}) \\
    \bottomrule
  \end{tabular}
\end{table}

\subsection{Urgency Growth and Beaconing}
Each active demand $d$ maintains an urgency $u_d$ that integrates linearly:
\begin{equation}
u_{d}^{t+1}=\min\!\big(u_{\max},\,u_d^t+u_{\mathrm{growth}}\big).
\end{equation}
The reseed into \eqref{eq:de} is
\begin{align}
S_{\mathrm{DE}}(x,t)
  &= \big(s_0+\alpha\,u_d^t + \mathbb{I}\{\text{beacon}(d)\}\cdot \text{beacon\text{-}bonus}\big)\,\notag\\
  &\qquad \times \mathbb{I}\{x=\text{loc}(d)\}, \label{eq:sde}
\end{align}

capped by \texttt{de-cap}. Beaconing triggers on first carrier contact; it deepens the local sink without changing the time-to-first-response.

\subsection{Motion, Separation, and Service}
Each carrier evaluates
\begin{equation}
\U(x)=\phide(x)-\kap\,\phihz(x)-0.15\,n_{\mathrm{here}}+\varepsilon, \label{eq:utility}
\end{equation}
where $\kap\!\ge\!0$, $n_{\mathrm{here}}$ discourages congestion, and $\varepsilon\!\sim\!\mathrm{Unif}(0,10^{-2})$ for tie-breaking. If another carrier is within radius $0.8$, they sidestep; with small probability they wander. A demand at $x$ with $m$ carriers within radius $r_s$ reduces work by $m\cdot\text{service\text{-}rate}$/tick; it is \emph{served} at zero remaining time or \emph{missed} after a deadline (creation$\!\to\!$now) controlled by \texttt{miss-cutoff}.

\newcommand{\1}[1]{\mathbb{I}\!\left\{#1\right\}}

\begin{algorithm}[t]
\caption{\textsc{X--Sycon} per–tick loop (no communication, local rules)}
\label{alg:loop}
\begin{algorithmic}[1]
\Require grid patches $P$ storing $\phi_{\mathrm{DE}},\phi_{\mathrm{HZ}},\texttt{blocked?}$; carriers $\mathcal C$; active demands $\mathcal D$.
\If{$t \ge \texttt{max\_ticks}$ \textbf{ or } $(\texttt{de\_served}+\texttt{de\_missed}) \ge \texttt{de\_quota}$}
    \State \textsc{Finalize}() and \textbf{halt}
\EndIf
\State \textsc{UpdateUrgency}$(\mathcal D)$  \Comment linear growth; clamp at $u_{\max}$
\State \textsc{UpdateFields}$(P,\mathcal D)$ \Comment decay, leak, reseed, diffuse
\State \textsc{EvolveEnvironment}$(P,\mathcal D)$ \Comment Bernoulli DE spawns; mean–reverting hazards
\State \textsc{MoveCarriers}$(\mathcal C,P)$ \Comment greedy ascent of $U$ with separation/wander
\State \textsc{ServiceAndComplete}$(\mathcal D,\mathcal C)$ \Comment first contact $\rightarrow$ beacon; completion/miss
\State $t \gets t+1$
\end{algorithmic}
\end{algorithm}

\begin{algorithm}[t]
\caption{\textsc{UpdateFields}: decay, reseed, diffuse}
\label{alg:fields}
\begin{algorithmic}[1]
\ForAll{$p\in P$}
  \State $\phi_{\mathrm{DE}}(p)\gets (1-\delta_{\mathrm{DE}})\,\phi_{\mathrm{DE}}(p)$;\quad
         $\phi_{\mathrm{HZ}}(p)\gets (1-\delta_{\mathrm{HZ}})\,\phi_{\mathrm{HZ}}(p)$
\EndFor
\If{$\texttt{leakDE}>0$}
  \ForAll{$p\in P$} \State $\phi_{\mathrm{DE}}(p)\gets \max\{0,\;\phi_{\mathrm{DE}}(p)-\texttt{leakDE}\}$ \EndFor
\EndIf
\ForAll{$p\in P$ with $\texttt{blocked?}=1$} \State $\phi_{\mathrm{HZ}}(p)\gets 12$ \Comment hazard reseed \EndFor
\ForAll{$d\in\mathcal D$ active}
  \State $s \gets s_0+\alpha\,u_d + \1{\textsc{Beacon}(d)}\cdot \texttt{beacon\_bonus}$
  \State at $p=\textsc{loc}(d)$: $\phi_{\mathrm{DE}}(p)\gets \min\{\phi_{\mathrm{DE}}(p)+s,\;\texttt{de\_cap}\}$
\EndFor
\State \textbf{Diffuse} $\phi_{\mathrm{DE}}$ with $D_{\mathrm{DE}}$; \textbf{Diffuse} $\phi_{\mathrm{HZ}}$ with $D_{\mathrm{HZ}}$ \Comment von Neumann conservative diffusion
\end{algorithmic}
\end{algorithm}

\begin{algorithm}[t]
\caption{\textsc{MoveCarriers}: local controller (per carrier $i$)}
\label{alg:move}
\begin{algorithmic}[1]
\ForAll{$i\in\mathcal C$}
  \If{$\exists j\neq i:\;\|p_i-p_j\|<0.8$} \Comment separation
     \State sidestep ($\pm 30^\circ$) and advance $1$ cell if free; increment energy; \textbf{continue}
  \EndIf
  \State $S \gets \{\,p'\in \mathcal N_4(p_i):\texttt{blocked?}(p')=0\,\}$ \Comment von Neumann neighbors
  \If{$S\neq\emptyset$}
     \State choose $p^\star \in \arg\max_{p'\in S}\;[\;\phi_{\mathrm{DE}}(p')-\kappa\,\phi_{\mathrm{HZ}}(p')-0.15\,n_{\text{here}}(p')+\varepsilon\;]$
     \State move one step toward $p^\star$ if free; increment energy
  \ElsIf{$\mathrm{Uniform}(0,1)<\texttt{wander\_prob}$}
     \State random turn ($\pm 45^\circ$ or $\pm 90^\circ$) and advance if free; increment energy
  \EndIf
\EndFor
\end{algorithmic}
\end{algorithm}

\subsection{Theory: Recruitment Radius and Service Bound}
Under a continuum approximation of \eqref{eq:de}–\eqref{eq:hz},
\begin{equation}
\partial_t \phi \approx D \nabla^2\phi - \lambda \phi + S,
\end{equation}
the Green’s function decays with characteristic length
\begin{equation}
\ell \approx \sqrt{D/\lambda}. \label{eq:length}
\end{equation}
A work-conservation bound gives sustainable throughput for $C$ carriers:
\begin{equation}
\mu_{\max} \lesssim \frac{C\cdot \text{service\text{-}rate}}{\E[B]}. \label{eq:bound}
\end{equation}

\subsection{Parameters, Sweeps, and Metrics}
\label{sec:params}

\begin{table}[!t]
  \centering\footnotesize
  \setlength{\tabcolsep}{5pt}\renewcommand{\arraystretch}{1.05}
  \caption{Fixed parameters}
  \label{tab:fixed}
  \begin{tabular}{@{}ll ll@{}}
    \toprule
    \textbf{Parameter} & \textbf{Value} & \textbf{Parameter} & \textbf{Value} \\
    \midrule
    diffDE  & 0.2  & decayDE & 0.02 \\
    leakDE  & 0.01 & de-cap  & 25   \\
    diffHZ  & 0.05 & decayHZ & 0.01 \\
    \bottomrule
  \end{tabular}
\end{table}

\begin{table}
  \centering\footnotesize
  \setlength{\tabcolsep}{5pt}\renewcommand{\arraystretch}{1.05}
  \caption{Urgency and service settings}
  \label{tab:urgency}
  \begin{tabular}{@{}ll ll@{}}
    \toprule
    \textbf{Parameter} & \textbf{Value} & \textbf{Parameter} & \textbf{Value} \\
    \midrule
    $u_{\mathrm{growth}}$ & 0.05 & $\alpha$   & 0.3 \\
    beacon-bonus          & 4    & $u_{\max}$ & 10  \\
    service-rate          & 1    & $r_s$      & 2   \\
    \bottomrule
  \end{tabular}
\end{table}

\begin{table}[!t]
  \centering\footnotesize
  \setlength{\tabcolsep}{5pt}\renewcommand{\arraystretch}{1.05}
  \caption{BehaviorSpace sweep ranges}
  \label{tab:sweeps}
  \begin{tabular}{@{}ll@{}}
    \toprule
    \textbf{Variable} & \textbf{Values} \\
    \midrule
    $C$                & \{10, 25, 50, 100\} \\
    $p_{\mathrm{new}}$ & \{0.02, 0.03, 0.04, 0.06\} \\
    hazard             & \{0.10, 0.15, 0.22, 0.28\} \\
    $\kappa$           & \{0.5, 0.8, 1.2, 1.6\} \\
    \bottomrule
  \end{tabular}
\end{table}

With $10$ seeds per setting, runs were stopped at $5{,}000$ ticks or $300$ tasks. We report the following:
(i) \emph{service throughput} $=\text{served}/\text{ticks}$,
(ii) \emph{miss rate} $=\text{missed}/\text{spawned}$,
(iii) mean TTFR and time in system (TiS),
(iv) \emph{energy per task}, and
(v) \emph{coverage entropy} using $16{\times}16$ bins:
\begin{equation}
H=-\sum_{i=1}^{256} p_i \ln p_i,\qquad 0\le H\le \ln 256\approx 5.545.
\end{equation}

\paragraph*{Operational definitions}
\emph{Time to first response (TTFR)} is the tick difference between a demand's creation and the first carrier contact (reported as $-1$ if no contact occurs).
\emph{Time in system (TiS)} is the tick difference between creation and completion/miss (this is the quantity our code logs as “latency”). Completion latency (first contact $\to$ completion) can be computed post hoc when $t_{\mathrm{first}}\neq -1$, but is not the default reporter.

\subsection{Implementation}
We implemented X--SYCON in NetLogo~6.4 with BehaviorSpace (\cref{fig:qual-snapshot}); logs capture per-DE events and per-run summaries. All figures in \cref{sec:results} are generated from these CSVs.
\paragraph*{Per-agent complexity and real-time budget}
Each carrier evaluates $\U(x)$ over its von\,Neumann neighborhood (four cells) with a one-sidestep rule and a constant-probability wander check. Per tick, the work per agent is $O(1)$ time and $O(1)$ memory, and the swarm step is $O(C)$ where $C$ is the carrier count. Field transport uses NetLogo's conservative diffusion/decay, which is linear in grid size ($33\times33$ here, fixed across experiments). Thus, wall-clock per tick scales linearly with $C$ at fixed map size; in practice our runs with $C\in\{10,100\}$ executed faster than real time on our test machine (specs in Appendix), though wall-clock time depends on hardware and logging. We report asymptotics and constant factors qualitatively here; the exact wall-clock depends on the interpreter speed and logging settings.
\paragraph*{Logging semantics and initialization details}
A missed task is marked as being \texttt{missed?=true} and removed like a completed one; global counters \texttt{de-served} and \texttt{de-missed} disambiguate outcomes. For reproducibility, we initialize with up to two demands (if under the active cap) and set initial hazards to approximately $\sim$60\% of the target blocked fraction to allow mean-reversion.

\section{Results}
\label{sec:results}

\begin{figure}[!t]
  \centering
  \includegraphics[width=\figw]{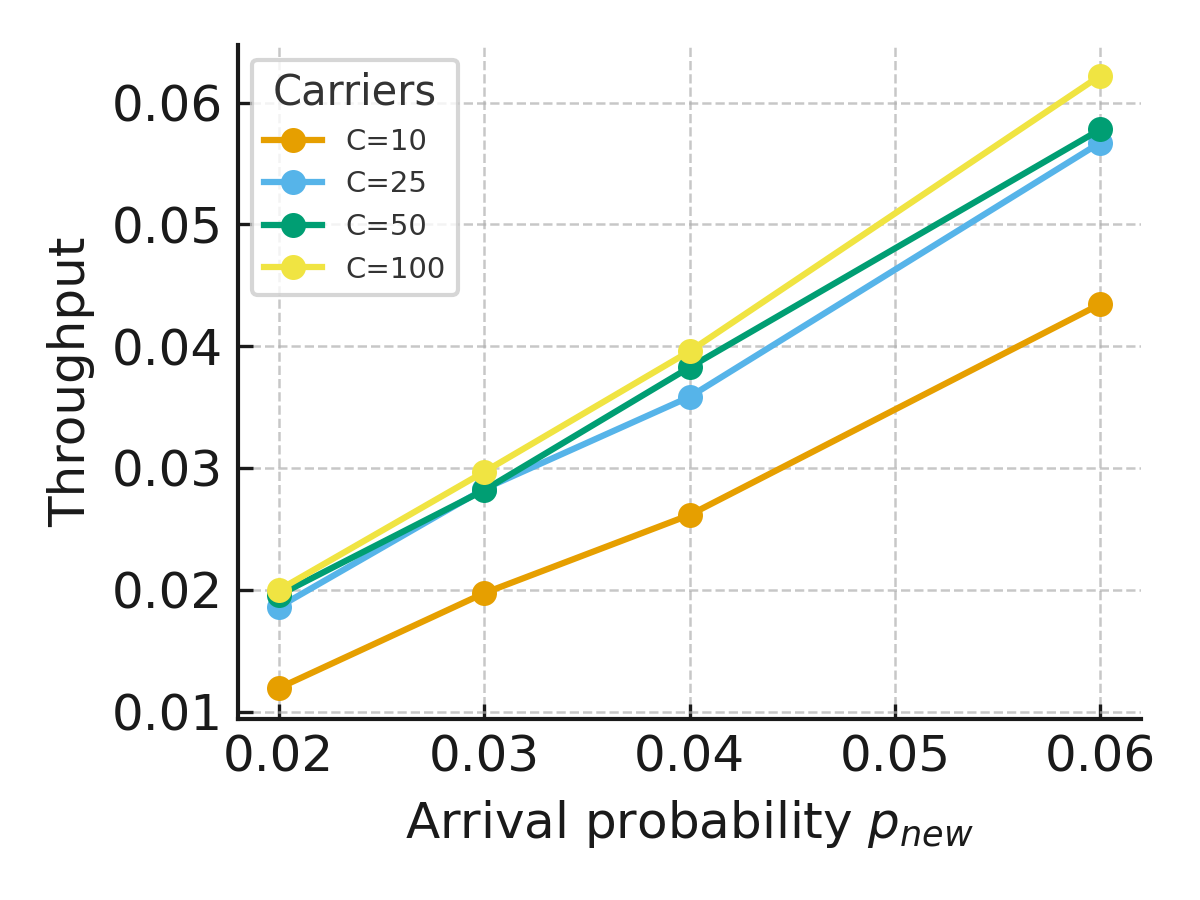}
  \caption{Throughput vs.\ arrival by carrier count (hazard 0.15, $\kappa{=}1.2$).}
  \label{fig:res-throughput}
\end{figure}

\begin{figure}[!t]
  \centering
  \includegraphics[width=\figw]{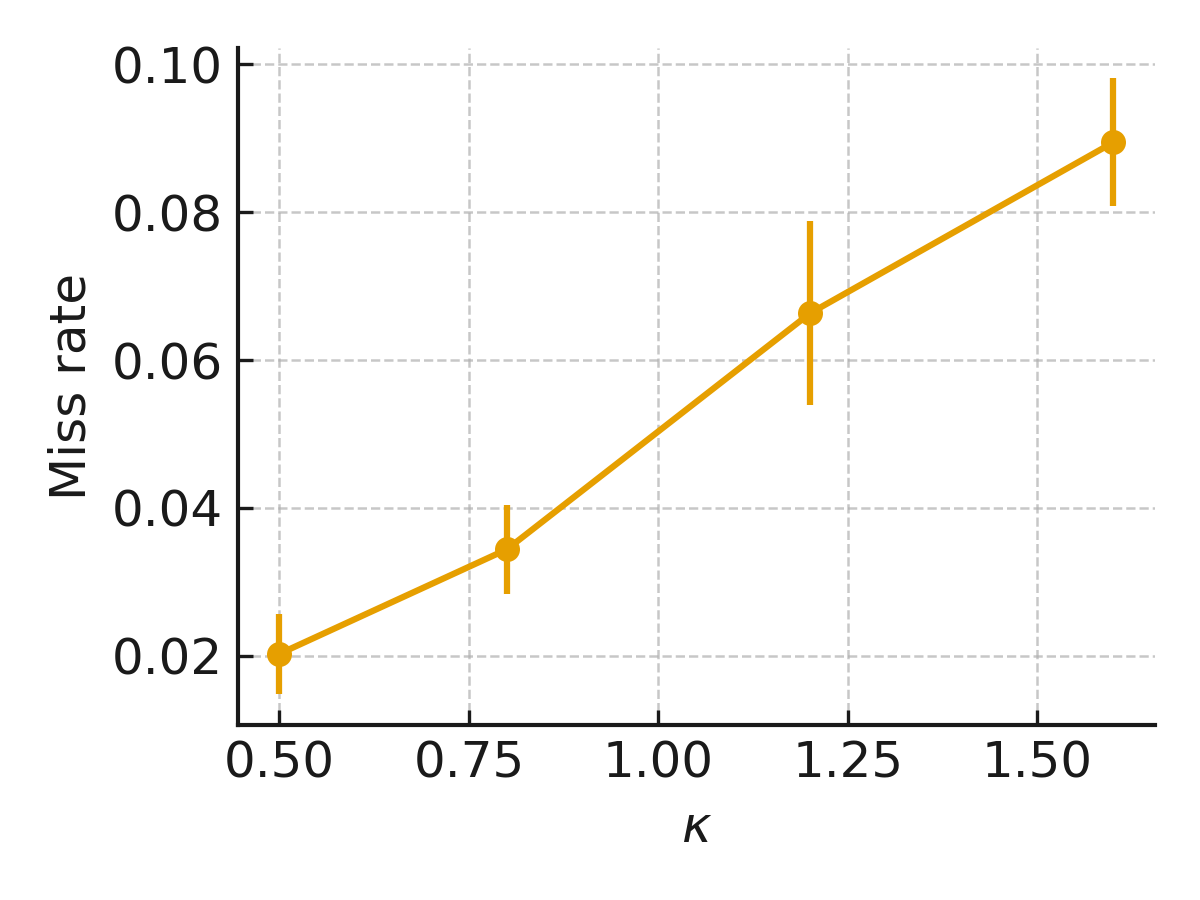}
  \caption{Miss rate vs.\ $\kappa$ at baseline (25 carriers, hazard 0.15, arrival 0.03).}
  \label{fig:res-kappa}
\end{figure}

We analyze \num{2560} NetLogo runs spanning:
\[
\begin{aligned}
C &\in \{10,25,50,100\}, \quad
p_{\mathrm{new}} \in \{0.02,0.03,0.04,0.06\},\\
\text{hazard} &\in \{0.10,0.15,0.22,0.28\}, \quad
\kappa \in \{0.5,0.8,1.2,1.6\}.
\end{aligned}
\]
Unless stated otherwise, the panels report the means across the seeds.

\subsection{Hazard sensitivity at a fixed operating point}
In the baseline slice ($C{=}25$, hazard $0.15$, $p_{\mathrm{new}}{=}0.03$), the miss rate increased with $\kappa$ (\cref{fig:res-kappa}). Penalizing the diffusive hazard halo too strongly weakens recruitment, raising misses and time-in-system (TiS) when blocked cells are uncrossable.

\subsection{Stability landscape in $(C,p_{\mathrm{new}})$}
For hazard $0.15$ at $\kappa{=}1.2$, the mean miss rate decreases with either more carriers or higher arrivals (\cref{fig:res-heat}). We hypothesize that increasing $C$ raises parallelism and service capacity, while increasing $p_{\mathrm{new}}$ sustains a stronger demand field $\phide$ which recruits more frequently. As a result, effective sinks can be deepened.

\subsection{Throughput scaling and saturation}
Throughput increases with arrival and, for our tested arrival rates, approaches the arrival ceiling; gains with additional carriers are sublinear (\cref{fig:res-throughput}). The Ohm's-law bound (\cref{eq:bound}) provides an ultimate ceiling at higher arrivals.

\subsection{Energy–reliability frontier}
Across all runs, energy per task and miss rate form a trade (\cref{fig:res-energy}): pushing the miss rate lower typically requires more motion/hover near sinks, raising energy. The frontier exposes a policy knob for tuning safety against budgets and quantifies an energy–reliability trade-off.

\subsection{Coverage structure}
\begin{figure}[!t]
  \centering
  \includegraphics[width=\figw]{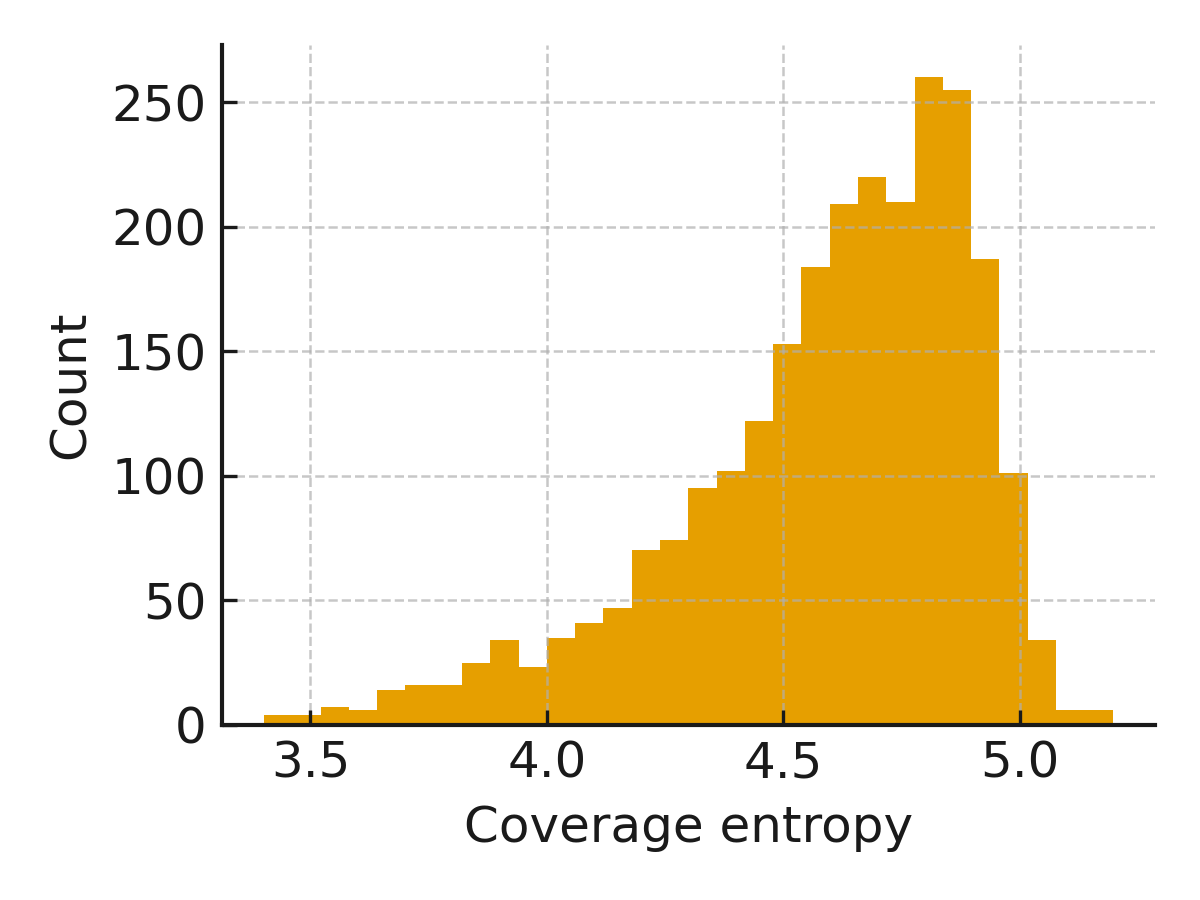}
  \caption{Coverage entropy distribution (16$\times$16 bins).}
  \label{fig:res-entropy}
\end{figure}

The cverage entropy concentrates well below the uniform bound $\ln 256\approx5.545$ (\cref{fig:res-entropy}). Agent trajectories exhibit lane-like patterns around obstacles, with exploration maintained via diffusion and random wander.

\subsection*{Findings}
\begin{figure}[!t]
  \centering
  \includegraphics[width=\figw]{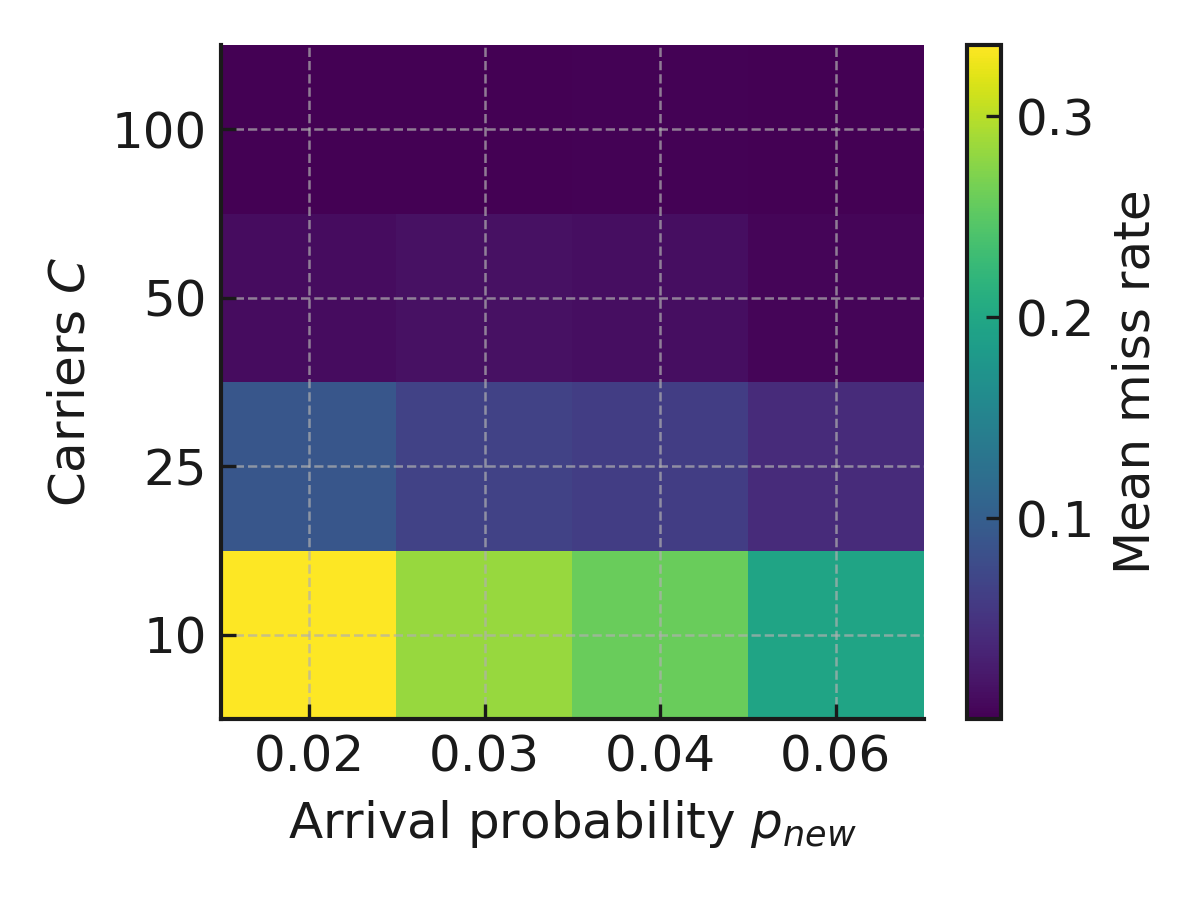}
  \caption{Mean miss rate over $(C,\,p_{\mathrm{new}})$ at hazard 0.15 and $\kappa{=}1.2$.}
  \label{fig:res-heat}
\end{figure}

\begin{figure}[!t]
  \centering
  \includegraphics[width=\figw]{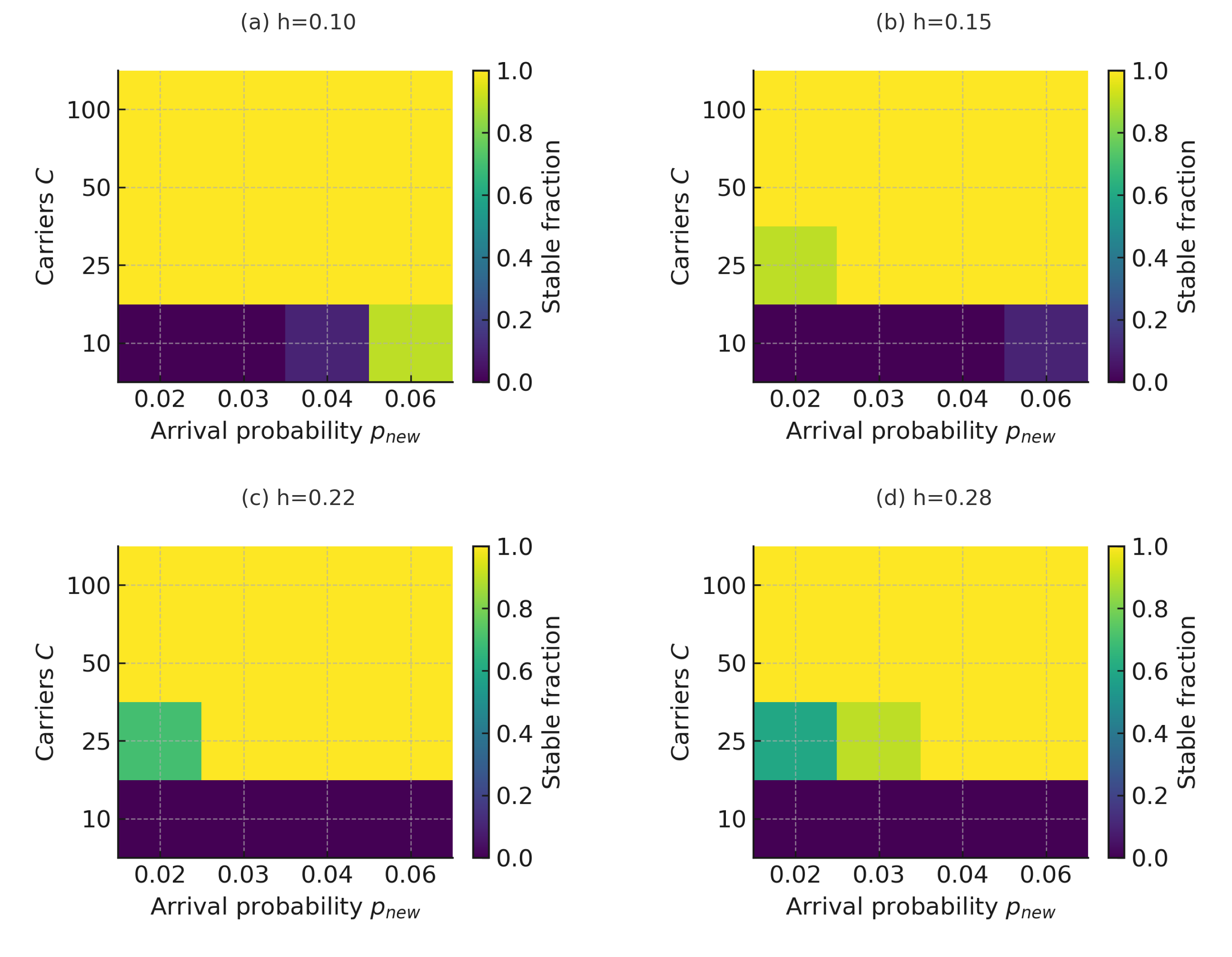} 
  \caption{Phase stability (fraction of seeds with miss rate $<0.15$) vs.\ carriers $C$ and arrival $p_{\mathrm{new}}$ at $\kappa{=}1.2$. Panel mapping: top–left $h{=}0.10$, top–right $h{=}0.15$, bottom–left $h{=}0.22$, bottom–right $h{=}0.28$.}
  \label{fig:res-phase}
\end{figure}

\noindent\textbf{F1:} Penalize hazards lightly when obstacles are hard. A large $\kappa$ over-penalizes the hazard halo and weakens the sink pull, increasing the number of misses (\cref{fig:res-kappa}).\\
\textbf{F2:} Throughput saturates sublinearly and reliability scales steeply. Capacity gains diminish with $C$, but miss rates decrease sharply (\cref{fig:res-throughput}).\\
\textbf{F3:} Stronger arrivals can reduce misses.
Higher $p_{\mathrm{new}}$ sustains $\phide$, increasing recruitment frequency; thus, beaconing is completed quickly (\cref{fig:res-heat}).\\
\textbf{F4:} Stable phases contract with hazards but are recoverable. 
More carriers or higher arrivals restored stability as hazard density increased (\cref{fig:res-phase}).
\section{Discussion}

\begin{figure}[!t]
  \centering
  \includegraphics[width=\figw]{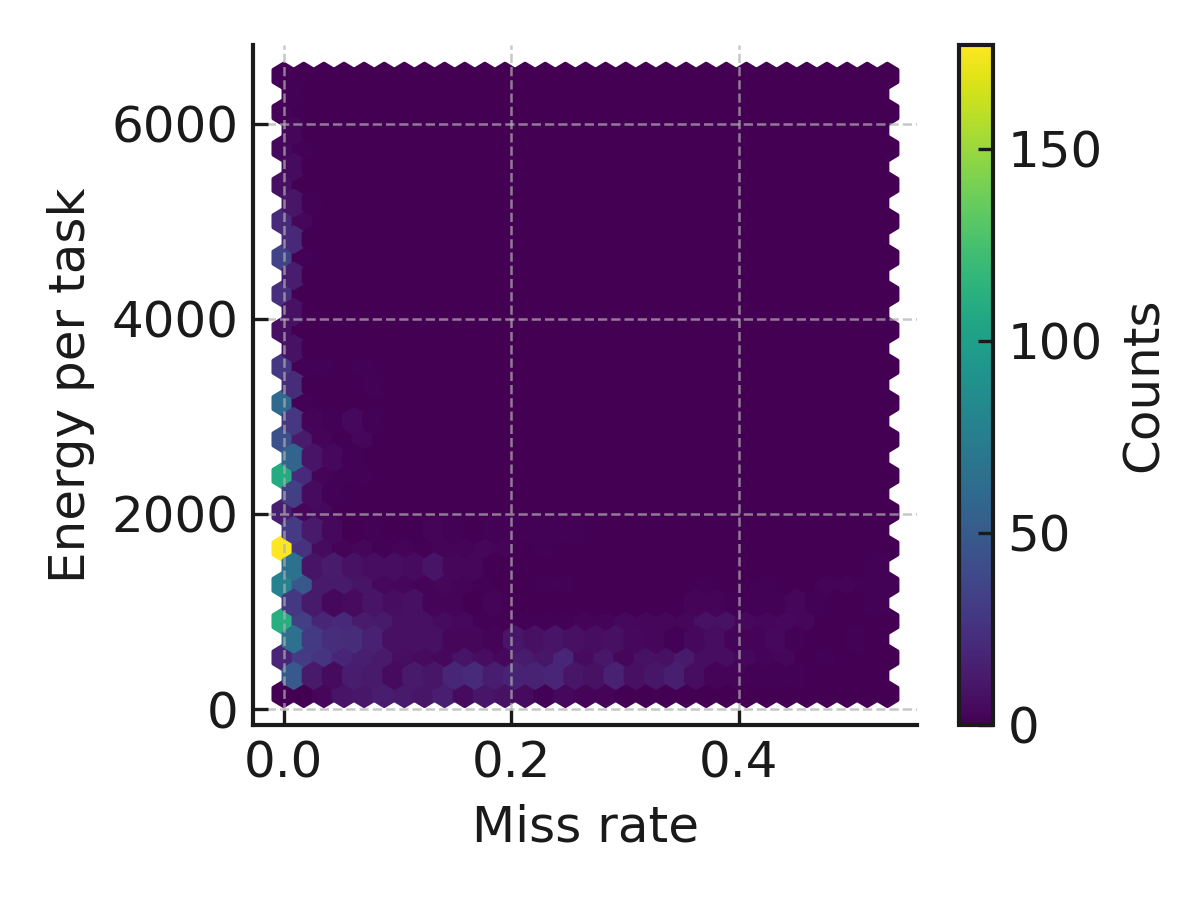}
  \caption{Energy per task vs.\ miss rate (all runs).}
  \label{fig:res-energy}
\end{figure}

In our simulations, the \emph{passive field dynamics} of X--SYCON achieved coordination without explicit communication in disaster-style scenarios. Three design lessons recur across our sweeps. First, when obstacles are already hard constraints, \emph{penalize hazards softly}: a large $\kap$ suppresses the diffusive halo that recruits help around viable corridors, raising misses. Second, \emph{scale reliability with carriers and scale capacity with sinks}. Throughput is arrival-limited in our sweeps (sublinear gains with $C$), and the work-conservation bound (\cref{eq:bound}) provides the ultimate ceiling at higher arrivals, but miss rates collapse quickly as carriers rise—an attractive operating mode for safety-critical response. Third, \emph{sink strength} was maintained. Higher $p_{\mathrm{new}}$ deepens $\phide$ and increases recruitment frequency; beaconing then reduces TiS (and qualitatively the completion interval) without affecting TTFR, which is equivalent to a temporary increase in the local demand source.

Conceptually, X--SYCON performs \emph{Distributed Passive Computation}: diffusion, decay, and reseeding implement a physical priority queue in which urgency integrates into potential, hazards shape feasible corridors, and multi-agent load sharing emerges from superposition. The hydraulic length scale $\ell$ chiefly governs the recruitment reach of $\phide$, whereas $\phihz$ steers routing via the penalty $\kappa$. The hydraulic length scale $\ell\!\approx\!\sqrt{D/\lambda}$ offers a knob linking controller design to environment: larger $\ell$ enlarges recruitment radii (faster response) but risks lane interference at a high hazard density. Practically, tuning begins with (i) pick $\kap$ at the smallest value that reliably avoids true hazards; (ii) target $\ell$ to the typical inter-incident spacing; (iii) add carriers until miss rates meet policy; (iv) use arrival-side levers (task admission / triage) to deepen sinks under stress.

Energy–reliability frontiers (\cref{fig:res-energy}) make trade-offs explicit: improved reliability costs motion/hover power near sinks. Because the trade-off was similar across hazard settings in our experiments, pre-computed policy curves may be feasible.

Finally, the qualitative \emph{field snapping} we observe—branching lanes and shared corridors—suggests an approach to mixed human–robot teaming: human responders could read heatmaps and place temporary ``beacons'' (virtual waypoints) to modulate flow without networked control. We leave formalizing this human-in-the-loop variant and learning $\ell$ from data to future work.

\section{Limitations and Ethics}

\paragraph*{Threats to validity}
\textit{Discretization.} The results use a $33{\times}33$ lattice and von\,Neumann neighborhoods; corridor percolation and boundary effects varied with domain size and obstacle morphology. 
\textit{Isotropy.} Diffusion and decay are isotropic and time-invariant; anisotropic transport (wind, slopes, and thermal plumes) would bias corridor formation. 
\textit{Perfect sensing.} Agents sample fields without noise or occlusion; modest noise is expected to blur $\nabla\U$ but not change qualitative trends, whereas severe occlusion reduces recruitment range. 
\textit{Homogeneity.} Carriers are identical; heterogeneous speeds or failure rates would widen latency and energy distributions. 
\textit{Energy accounting.} Step-based, normalized costs omit battery dynamics and payload penalties, and deployments should substitute task-independent baselines.
\textit{TTFR. } Our TTFR reporter returns $-1$ when no first contact occurs, and TiS remains defined in those cases (creation$\to$miss).

In our study, sensing, actuation, and failure were abstracted. Sensors are perfect, carriers are homogeneous, and motion is grid-constrained with unit steps; we do not model slip, drift, occlusions, or communications interference beyond the deliberate omission of messaging. Diffusion and decay are isotropic and time-invariant; anisotropic transport (e.g., wind, slopes, thermal plumes) alters corridor formation. The world is $33{\times}33$; while the controller is scale-free, boundary effects and corridor percolation change with domain size and obstacle morphology.

Energy accounting is step-based and normalized; we do not include the battery dynamics, charging, or payload costs. When no tasks are served, our reporter safeguards against division by zero, biasing energy-per-task downward. Real deployments should use task-independent energy baselines. Beaconing is an idealized local reseed. In hardware, the equivalent (e.g., a strobe or radio-free marker) may be rate-limited or intermittent.

Ethically, X--SYCON is intended for safety and infrastructure protection purposes. The very features that make it resilient (no communication, decentralized) also make monitoring more difficult. Deployments should log local decisions and maintain geographic exclusion around known hazards to avoid emergent harm. Human oversight remains critical in triage and escalation. We avoided training on sensitive data; all code and synthetic data were released for reproducibility.

\section{Conclusion}

We presented X--SYCON, a xylem-inspired controller in which coordination emerges from passive fields rather than messages. In dynamic, partially blocked worlds, the greedy ascent of $U=\phide-\kap\phihz$ with light anti-congestion achieves low miss rates and stable throughput. The hydraulic length scale is related to recruitment, and the work-conservation bound was consistent with observed sublinear capacity. Empirically, soft hazard penalties, stronger sinks, and modest carrier counts yield reliable service, thereby exposing a stable energy–reliability frontier. We frame this as \emph{Distributed Passive Computation and Control}, which is an approach evaluated under communication-denied assumptions. Future work could include anisotropic transport, noise sensing, scale-up on irregular maps, and human-placed beacons. 
 
\section*{Declarations}

\noindent\textbf{Competing interests.}
The authors declare \emph{no competing interests}.

\medskip
\noindent\textbf{Funding.}
This research \emph{received no specific grant} from any funding agency in the public, commercial, or not-for-profit sectors.

\medskip
\noindent\textbf{Data availability.}
All data and code supporting the findings of this study are publicly available at
\url{https://github.com/arthurbaek/biovascular-swarm-robotics}.
The exact version used to produce the figures is the tagged GitHub release \texttt{v1.0.0}.
(If an archival snapshot is later deposited, the DOI will be added in a revised version.)

\medskip
\noindent\textbf{Ethics approval.}
This study did not involve human participants, animal experiments, or field studies requiring institutional review.
No ethical approval was required.

\balance
\bibliographystyle{IEEEtran}
\bibliography{references}

\appendices
\section{Supplementary Material}
Complete NetLogo code and BehaviorSpace grids are available in the project repository. For double\textendash blind review, we provided an anonymized artifact.

\section*{Artifacts}
Code/data: \href{https://github.com/arthurbaek/biovascular-swarm-robotics}{github.com/arthurbaek/biovascular-swarm-robotics}.
(For double-blind submission, replace with an anonymized link.)

\end{document}